# From internal models toward metacognitive AI

Mitsuo Kawato and Aurelio Cortese
ATR Brain Information Communication Research Group,
Computational Neuroscience Laboratory, Hikaridai, Kyoto Japan 619-0288

**Abstract**
In several papers published in *Biological Cybernetics* in the 1980s and 1990s, Kawato and colleagues proposed computational models explaining how internal models are acquired in the cerebellum. These models were later supported by neurophysiological experiments using monkeys and neuroimaging experiments involving humans. These early studies influenced neuroscience from basic, sensory-motor control to higher cognitive functions. One of the most perplexing enigmas related to internal models is to understand the neural mechanisms that enable animals to learn large-dimensional problems with so few trials. Consciousness and metacognition the ability to monitor one's own thoughts, may be part of the solution to this enigma. Based on literature reviews of the past 20 years, here we propose a computational neuroscience model of metacognition. The model comprises a modular hierarchical reinforcement-learning architecture of parallel and layered, generative-inverse model pairs. In the prefrontal cortex, a distributed executive network called the "cognitive reality monitoring network" (CRMN) orchestrates conscious involvement of generative-inverse model pairs in perception and action. Based on mismatches between computations by generative and inverse models, as well as reward prediction errors, CRMN computes a "responsibility signal" that gates selection and learning of pairs in perception, action, and reinforcement learning. A high responsibility signal is given to the pairs that best capture the external world, that are competent in movements (small mismatch), and that are capable of reinforcement learning (small reward-prediction error). CRMN selects pairs with higher responsibility signals as objects of metacognition, and consciousness is determined by the entropy of responsibility signals across all pairs. This model could lead to new-generation AI, which exhibits metacognition, consciousness, dimension reduction, selection of modules and corresponding representations, and learning from small samples. We may also develop a new scientific paradigm that enables the causal study of consciousness by combining CRMN and decoded neurofeedback.

**Acknowledgements**
Hakwan Lau, Jorge Morales, Matthias Michel, Samuel Gershman, Megan Peters and JD Knotts read an earlier version of the manuscript. We thank them for valuable comments and questions that were incorporated in the current manuscript as much as possible. This study was supported by Innovative Science and Technology Initiative for Security Grant Number JPJ004596, Acquisition, Technology & Logistics Agency (ATLA), Japan. This study was also supported by the Grant Number JP21dm0307008, Japan Agency for Medical Research and Development (AMED) and JPMJER1801, ERATO, Japan Science and Technology Agency (JST).

# (1) Introduction

Internal models are neural networks in the brain that simulate dynamics of some aspects of the external world. In the context of sensory-motor control, two kinds of internal models of a controlled object, a forward model and an inverse model, are possible to simulate objects of motor control. Examples of controlled objects are a robotic manipulator and a human body. A controlled object receives motor commands, such as joint torques and muscle tensions, and outputs a movement trajectory, in the form of joint angles and muscle lengths. A forward model is an internal model of the controlled object with the same "from-input-to-output" direction as the controlled object. Thus, a forward model receives a motor command and predicts a movement trajectory. In a neuroscience context, a forward model receives an efference copy of a motor command, and predicts the resulting sensory feedback caused by executed movements. An inverse model is also a model of the controlled object, but with the opposite "from-input-to-output" direction; thus, it receives a desired trajectory as input and computes a motor command necessary to realize the desired trajectory. Connected in tandem, the inverse model and the forward model become an identity function, which is the reason that the inverse model can compute the necessary motor command, and may serve as an ideal feedforward controller.

Kawato and colleagues proposed that lateral and medial parts of the cerebellum acquire inverse and forward models of controlled objects through motor learning, and that they are hierarchically arranged (Kawato et al., 1987; Kawato and Gomi, 1992). Forward internal models are incorporated within internal feedback control loops, and inverse models are placed as feedforward controllers on top of the feedback loops. Forward models may also be utilized for optimal control of either kinematic or dynamic optimization objectives (Kawato 1999; Uno et al., 1989; Todorov and Jordan 2002). Supervised learning of forward models is straightforward because sensory feedback furnishes teaching signals in learning. However, learning of inverse models is difficult because we cannot assume the presence of "teaching signals", i.e., ideal motor commands in the brain. Feedback-error learning of inverse models postulated that feedback motor commands generated by either internal or external feedback loops could be used as approximate error signals for training inverse models (Kawato et al., 1987). Mathematical proofs of its stability and convergence were developed (Nakanishi and Schaal, 2004), and robotic applications demonstrated its practical utility (Miyamoto et al., 1988; Atkeson et al., 2000). Various experimental studies supported cerebellar internal models, especially inverse models, and its special case of feedback-error learning (Yamamoto et al., 2007). They include recording of simple spikes and complex spikes of monkey Purkinje cells during ocular following responses (Shidara et al., 1993; Kawato 1999; Kobayashi et al., 1998; Gomi et al., 1998; Yamamoto et al., 2002), and humans learning a new tool (fMRI study) (Imamizu et al. 2000).

In sensory-related cortices, fast visual computation by forward and inverse optics models was proposed (Kawato et al., 1993), motivated by Grossberg and Mingolla (1985) and Mumford (1992). Here, we use the word "optics" as the image generation process from properties of the three-dimensional external world including surface properties, object shapes, and light sources. The inverse optics model infers these latent variables related to the external world from visual images. The forward-inverse optics model explains very fast computation in the human visual system, while solving the inverse-optics or vision problem (Marr 1982; Poggio et al., 1985). Solving complicated nonlinear inverse problems usually requires many iterative computations, which is incompatible with human studies showing fast visual processing (Thorpe et al., 1996). The areas higher in the hierarchy of sensory cortices were assumed to represent the external world more abstractly, and those lower in the hierarchy, rawer representations of the external world. Feedback neural connections from higher to the lower visual area provide a forward optics model, or a generative model from latent variables to image data in recent terminology. A forward optics (generative) model reconstructs the rawer representation (e.g., sensory signals) from the more abstract representation (e.g., latent variable) of the external world. In contrast, feedforward neural connections from the lower area to the higher area are assumed to provide an inverse optics model, or an inference model in recent terminology, in other words, analytic one-shot computation estimating higher-order representations from lower-order representations. Because the inverse optics model provides an approximate solution to the inverse problem (vision) using a one-shot computation, the whole computation of vision can be fast. The forward optics model, on the other hand, guarantees accurate and stable solutions using iterative computations. Recurrent computations between the forward and inverse optics model were laid out in the laminar structures within hierarchical sensory cortices (Fig 1B of Kawato et al., 1993, also see Fig. 3a). Errors between the two models were proposed to be sent to higher cortices again, while filtered by the inverse optics model. This scheme was named "predictive coding" by Rao and Ballard (1998) and Friston and Kiebel (2009).

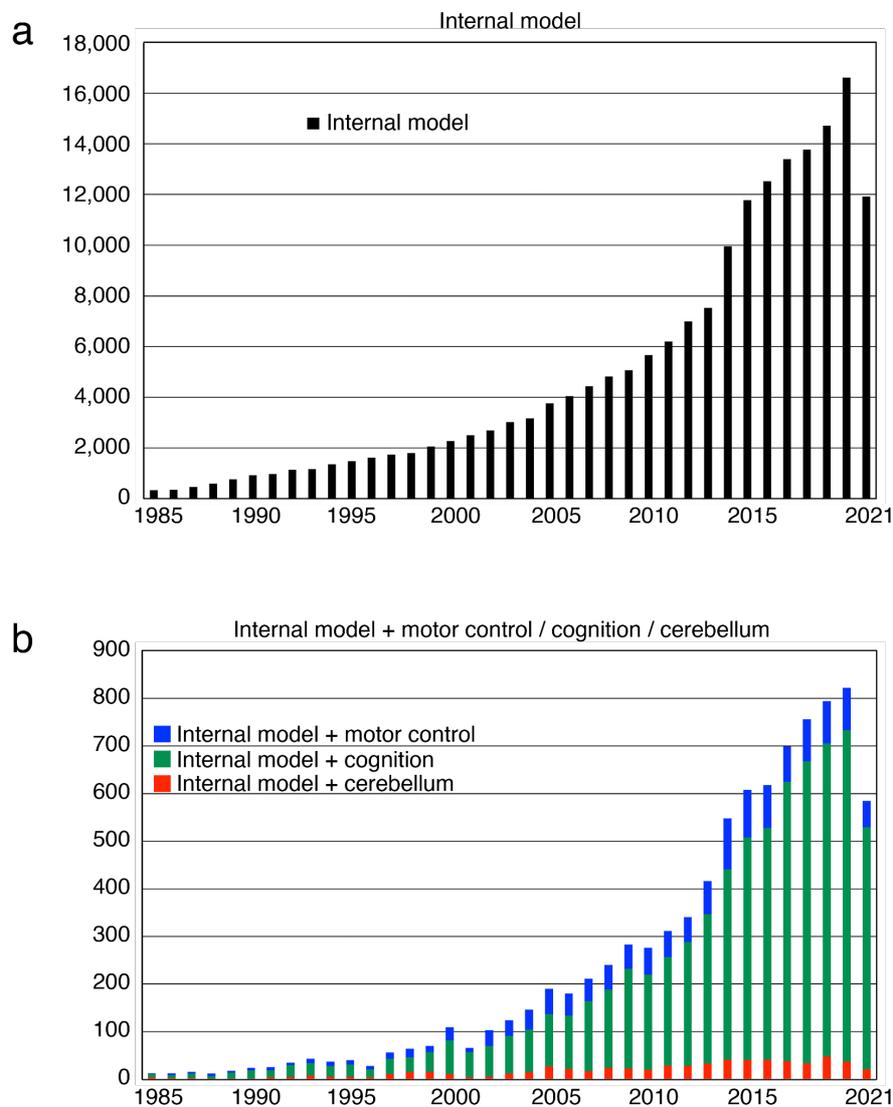

Fig. 1 **a.** Numbers of PMC papers published each year, as shown by PubMed searches for the keyword "internal model". The ordinate shows the number of published papers and the abscissa shows the year of publication. **b** Numbers of publications per year with combinations of "internal model" AND "motor control" (blue), "cognition" (green), or "cerebellum" (red). The search was conducted on 4 August 2021.

These early models were incorporated in subsequent studies and influenced sensory-motor control (Ito 2008; Shadmehr et al., 2010; Wolpert et al., 1998; Wolpert and Kawato 1998) and perceptual studies (Friston 2005, 2010; Friston et al., 2006; Kawato 1997; Lee and Mumford 2003; Olshausen and Field, 1996; Rao and Ballard, 1998). The original papers and related reviews on internal models were cited several thousand times each (Kawato 1999; Kawato et al., 1987; Wolpert et al., 1998). Figure 1a shows a rapid increase of the number of publications with the keywords "internal model", and 1b shows those with "internal model" AND "motor control", "cognition", or "cerebellum." The rapid increase was especially marked for "internal model" and "cognition". In the following sections, we discuss one of the remaining fundamental problems with internal models: how internal models of large, complicated objects can be learned with a small number of trials. Finally, based on the proposed model of metacognition, we speculate how we can develop a new consciousness research paradigm: causal study of consciousness.

## (2) Literature on learning from small samples

We review several lines of studies relevant to the topic of learning from small samples, where appropriate generalization can be achieved while utilizing limited training experiences. Generalization error is defined as the error in unseen test data (validation set), used as an objective performance measure of generalization for a learning algorithm. According to mathematical theories of learning, the generalization error can be estimated by several factors, including degrees of freedom of the learning system and the amount of data used for training. A classical evaluation provides that the generalization error is given as the degrees of freedom divided by twice the number of training samples (Watanabe 2009). Based on this evaluation, we know that a learning system with a small number of parameters can generalize even with a small number of training samples. More recent estimations, motivated by success of deep learning, provided a much weaker constraint on the number of necessary training samples (Suzuki 2018). However, these formulas are asymptotic estimates, and common sense suggests that we need hundreds of thousands of training samples for a learning machine with hundreds of thousands of learning parameters, e.g., the number of modifiable synapses. Human brains contain $10^{14}$ synapses, and if they require a comparable number of training samples, life expectancy of organisms is not sufficient to account for their capacity to learn. Even for a single cerebellar microzone, there are at least 10,000 Purkinje cells and $\geq 10^9$ plastic synapses. If one training sample is collected every second, this requires more than 30 years of training. A human body possesses at least ten million muscle fibers and millions of motor neurons; thus, motor learning problems are huge. However, animals can learn new movements, skills, and tasks within a few hundred trials, and even learn to avoid ingesting a toxin after just one trial in the case of taste aversion (Nikolaus et al., 1983, Roper & Redston 1987). Previous studies have offered several possible mechanisms for the remarkable ability of animals to learn large-scale problems from small samples. While discrimination between two stimuli can be learned with a single experience, the problem is also computationally simpler than learning an internal model. The latter is more difficult because it involves learning some dynamic aspect of the world, and as such, requires a much higher number of units and parameters. Here we review the following factors: modular and/or hierarchical structures, feature selection and/or dimensional reduction, and metacognition and/or consciousness.

The basic idea behind the modular architecture is to divide and conquer. By partitioning a huge problem into many small-task pieces, each learning module can deal with a tractably small piece. Modular neural-network models started with the pioneering "mixture-of-experts (MoEXP)" model (Jacobs et al., 1993). The MoEXP model was extended to the MOSAIC model, which contains both forward and inverse models (Wolpert and Kawato 1998; Haruno et al., 2001), and was further developed into reinforcement learning MOSAIC (RL-MOSAIC; Doya et al., 2002; Sugimoto et al., 2012a, 2012b). Hierarchy is another architecture that is based on the "divide and conquer" strategy. In the upper hierarchy, dimension reduction of the task is possible with coarse-grained representations of states and actions, while in the lower hierarchy, the task space is partitioned into small subregions in which solving the task is tractable. Hierarchical reinforcement learning was one of the strong theoretical fields with this computational objective (Wiering and Schmidhuber 1997; Parr et al., 1997), and Samejima and colleagues combined hierarchical reinforcement learning with multiple internal models (Samejima et al., 2003, 2006; Kawato and Samejima 2007). In the neuroscience of motor control, hierarchical models (Kawato et al., 1987) and uniform and flat models (Todorov and Jordan 2002) were proposed, and there have been oscillations back (Scott 2004) and forth (Franklin et al., 2008; Osu et al., 2015; Babič et al., 2016; Ikegami et al., 2021). In robotics and artificial intelligence, hierarchical reinforcement learning has been explored for almost 20 years (Morimoto and Doya 2001), and recently it has regained popularity (Sugimoto et al., 2012a; Merel et al., 2019). Kawato and colleagues proposed a cerebellar hierarchical reinforcement learning model based on these previous theoretical models, explaining recent experimental findings in the cerebellum (Kawato et al., 2020).

Finally, several researchers independently proposed that higher cognitive functions, especially consciousness (Bengio 2017) and metacognition (Cortese et al., 2019), serve very important functions in feature selection and dimension reduction, which are beneficial for learning from small samples in hierarchical structures of the brain. Consciousness and metacognition are in fact intimately related, and might even share common neural mechanisms (Brown et al., 2019, Morales & Lau 2021). Of interest here is that both consciousness and metacognition could be linked to higher-order representations; reflection, or re-representations of first-order sensory representations (Lau & Rosenthal 2011, Brown et al., 2019). Such higher-order representations are presumed to be abstract and low-dimensional (in content space), and are essential to feature selection and dimension reduction (Cortese et al., 2019, Fleming 2020). Recent theoretical work has discussed consciousness as a system invented for the brain's need to constantly discriminate what is internally generated from what is a true representation of the external world -- an internal mechanism of perceptual reality monitoring (Lau 2019). In similar fashion, Gershman (2019) suggested that generative adversarial networks (GAN, Goodfellow et al., 2014) provide a striking analogy for how the brain operates in metacognition and consciousness. The key idea in the context of consciousness is a discriminator of GAN between true and internally generated. The discriminator maps onto the higher-order representations, which are effectively belief states about lower-level representations.

## (3) Metacognition accelerates reinforcement learning

It is always very difficult to estimate how much genetic information is utilized when animals learn a huge dimensional problem from a small sample. For example, foals stand immediately after birth, and this capability must come largely from genetic hardwiring of motor neural circuits, but such estimation is difficult in neuroscience, for example, when humans learn to use new tools. As the first step to identify possible neural mechanisms of learning from small samples, we need to show that animals (humans) can learn big problems from small samples when genetic information or prior knowledge is *not* available.

Cortese and colleagues achieved this seemingly very difficult task by arbitrarily separating brain states into two domains, using a binary decoding technique. They prepared a reinforcement learning task in which participants had no clue about the reinforcement learning state (Cortese et al., 2020). This was made possible by extending the fMRI decoded neurofeedback method (Shibata et al., 2011, 2018; Watanabe et al., 2017), and by constructing a novel and innovative reinforcement-learning task. Multi-voxel decoding was used to separate the brain state into two domains, each associated with an optimal action that, if selected, would lead to high probability of reward. The decoding utilized about 200 voxels, but participants were unaware of which brain area was targeted for decoding and how these voxels were selected. Furthermore, fMRI BOLD signals used for this decoding were measured during inter-trial intervals of the reinforcement-learning task, during which there was no task for participants nor stimulus given. Consequently, participants had no prior information about the reinforcement learning state; thus, genetic information could not help to solve the reinforcement-learning task. Nevertheless, to our surprise, participants learned to select optimal actions within several hundred trials in just two days of the three-day experiment. This study clearly demonstrated that humans can learn gigantic problems (equivalent to ~10,000 voxels, and possibly $10^{14}$ synapses) from small samples, independent of genetic information. Experimental results obtained from the fMRI data demonstrated that the search space for the reinforcement-learning state started from the whole brain, and then rapidly shrank to very limited regions of the brain, including the basal ganglia and the prefrontal cortex (PFC). Metacognition proved important for this learning from small samples, based on the following three findings. First, participants with better metacognitive capability solved reinforcement learning more effectively, i.e., were more likely to select the optimal option; hence, they obtained larger rewards. Here, metacognitive capability is defined according to how well each participant can estimate correctness of their perceptual decisions on motion directions by their subjective confidence ratings. Second, when the confidence rating was high, then optimal choices as well as smaller reward prediction errors were observed more often. That is, there was information coupling between the confidence and the reward prediction errors. Finally, as learning progressed, the above information coupling between confidence decoded in the PFC and reward prediction errors decoded in the basal ganglia became stronger. That is, for later learning stages, when decoded confidence was greater, the decoded reward prediction error was smaller. The functional relationship between the two brain regions became stronger and stronger during three days of learning (Cortese et al., 2020).

In the next section, we propose a computational neuroscience model of metacognition. This model basically reproduces the aforementioned experimental data, and leads to next generation artificial intelligence with metacognition, consciousness, and learning from small samples. This computational model is an expansion and integration of the lines of research that were introduced in the previous section. First, the metacognition model contains multiple generative-inverse models. So, it is a natural extension of internal model theories (Kawato 1997, 1999; Kawato et al., 1993, 1987, 1992), forward-inverse optics models (Kawato et al., 1993), the MoEXP model (Jacobs et al., 1993), and MOSAIC (Wolpert and Kawato 1998; Haruno et al., 2001). Second, it is a modular-hierarchical reinforcement learning model; thus, it is an extension of RL-MOSAIC (Doya et al., 2002; Sugimoto et al., 2012a, 2012b). Third, the metacognition model was inspired by the hypothesis of perceptual reality monitoring (Lau 2019), as well as the model of consciousness by Gershman, based on a generative adversarial network (GAN) (Goodfellow et al., 2014; Gershman 2019).

## (4) A computational model of metacognition

The proposed model constitutes a hierarchical modular structure of the cerebral cortex. Within a given module at a certain hierarchy, a pair of generative and inverse models constitute an element. Pairs of conjugate models are arranged in hierarchy, as well as in parallel (Figs. 2 and 3a). In sensory related cortices (Fig. 2a), feedback neural connections from the higher area to the lower area provide a forward optics model (Forward: $g$), in other words, a generative model of the rawer representation (Eqs. 1, 2). In contrast, feedforward neural connections from the lower area to the higher area provide an inverse optics model (Inverse: $f$), in other words, an analytical, one-shot computation estimating higher-order representations (Eqs. 3, 4) (Kawato et al., 1993). In this paper we use deterministic formulas for simplicity, but one can develop the corresponding stochastic formulas (Friston 2010; Gershman 2019).

$$\hat{x}_L = g_L(x_H, \hat{x}_H) \tag{1}$$
$$\hat{x}_H = g_H(x_{HH}, \hat{x}_{HH}) \tag{2}$$
$$x_H = f_L(x_L, \hat{x}_L) \tag{3}$$
$$x_{HH} = f_H(x_H, \hat{x}_H) \tag{4}$$

Here, *LL, L, H, HH* denote one level lower than lower in hierarchy, lower level in hierarchy, higher level in hierarchy, higher than higher level in hierarchy, respectively. Within a cortical region, top-down computational outcomes from a generative model $\hat{x}$ and bottom-up computational outcomes from an inverse model $x$ are compared, and the mismatch between the conjugate-pair estimates is computed as in Eqs. 5,6.

$$\| \hat{x}_H - x_H \| \tag{5}$$
$$\| \hat{x}_L - x_L \| \tag{6}$$

This mismatch may originate from several factors. The first and most obvious is that the brain region in a hierarchy does not match the relevant aspect of the external world. If animals need to discriminate between their prey and predators based only on odors in total darkness and from remote distances, then visual or somatosensory modules would have large mismatches. This is because top-down and bottom-up computations largely diverge in these irrelevant sensory modalities. Secondly, large prediction errors are induced by inaccurate predictive, forward, and generative models when the models are not learned well. The third involves computation errors by inappropriate inverse models when they are poorly suited to a specific perceptual domain. In parallel with this mismatch error between paired models, the reward prediction error $\delta_{ik}$ utilizing the representation of that *(i,k)* cortical region is computed through communication with the basal ganglia (Fig. 3a). Here, *i* represents the number of modalities of modules, and *k* represents the depth in hierarchy and is designated as *LL, L, H, HH* in eqs. (1)~(6). For all conjugate-model pairs, mismatch errors and reward prediction errors are computed, and their weighted summation is used by the cognitive reality monitoring network (CRMN) in the PFC (Fig. 3a).

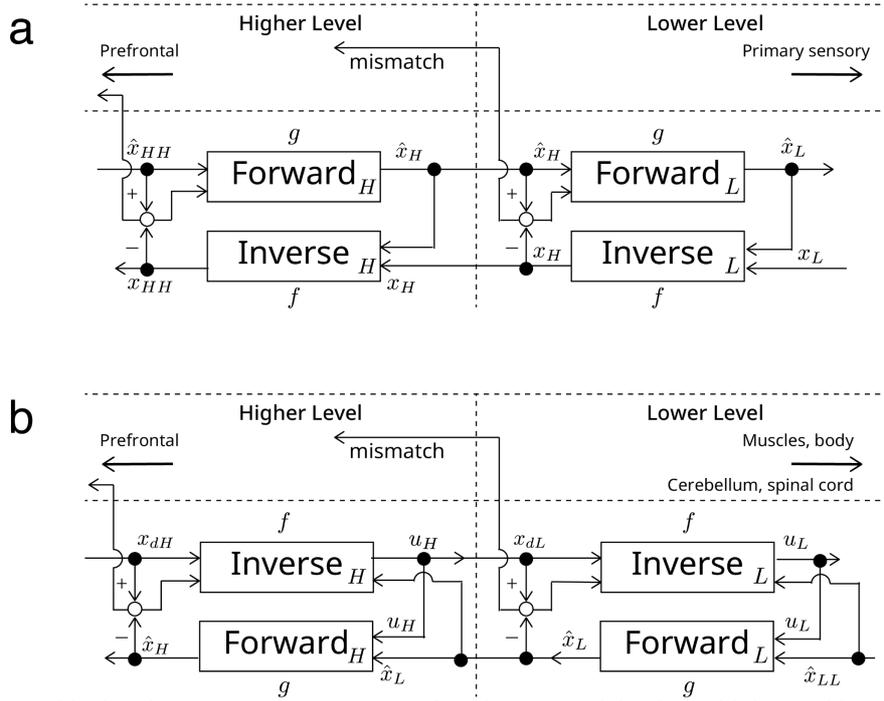

Fig. 2 **a**) Hierarchical and recurrent arrangements of conjugate-model pairs in higher and lower levels of sensory cortices. Forward and inverse models in the pair are a generative, forward optics model and its inverse in the case of vision. The mismatches between forward and inverse computations are calculated (open circles in the figure), and are used as inputs to forward models, as well as sent to the prefrontal cortex. $x$: A representation of the external world computed by feedforward one-shot, and analytical, inverse models (bottom up), $\hat{x}$: representation of the external world computed by feedback, iterative, generative, forward models (Top down); $H$: higher level in hierarchy; $L$: lower level in hierarchy; $HH$: higher than higher level in hierarchy. **b**) Hierarchical and recurrent arrangements of conjugate-model pairs in higher and lower levels of sensory-motor cortices. Forward and inverse models in the pair are a predictive, forward model and an inverse model of a controlled object at that level of representation. The mismatches between forward and inverse computations are calculated (open circles in the figure), and are used as inputs to inverse models, as well as sent to the prefrontal cortex. $x$: state; $\hat{x}$: predicted state by forward model; $x_d$: desired state; $u$: motor command; $H$: higher level in hierarchy; $L$: lower level in hierarchy; $LL$: one level lower than lower

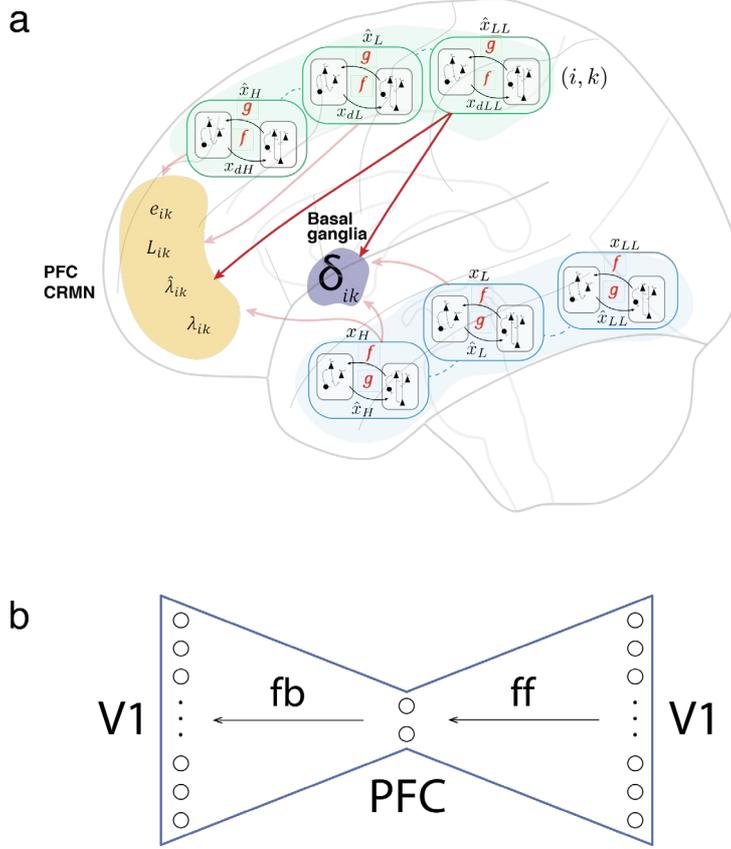

Fig. 3. **a** Whole brain parallel, hierarchical structure, having loop communications with the basal ganglia. Neural circuits within laminar structures of cerebral cortices are redrawn 12 times according to those in Fig. 1B of Kawato et al. (1993). Each hierarchy in each modality contains both forward and inverse models. The upper hierarchy represents motor cortices and the lower hierarchy represents sensory cortices. Note that forward models and inverse models are bottom-up (feedforward) and top-down (feedback) directions in the motor cortices, but they are reversed in the sensory cortices. CRMN in the PFC contains cognitive prediction errors $e_{ik}$, likelihoods $L_{ik}$, responsibility signals $\lambda_{ik}$ and their priors $\hat{\lambda}_{ik}$. The basal ganglia compute reward prediction errors $\delta_{ik}$ for all modality and hierarchy $(i, k)$. Here, $i$ represents the modality, such as vision, audition, somatosenses. $k$ represents the level in hierarchy, and corresponds to *LL, L, H, HH*. **b.** An autoencoder network when feedforward (ff) and feedback (fb) neural connections in **a** are unfolded in the right and left sides of the PFC, respectively. Here, for simplicity, the schematic representation only depicts V1 (early visual cortex), but the same mapping applies to any sensory or motor area. $x$: state; $\hat{x}$: predicted state by forward model; $x_d$: desired state; $u$: motor command; *H*: higher level in hierarchy; *L*: lower level in hierarchy; *LL*: one level lower than lower. $e_{ik}$, $L_{ik}$, $\lambda_{ik}$, $\hat{\lambda}_{ik}$, $\delta_{ik}$: cognitive error signal, likelihood, responsibility signal, responsibility-signal prior, and reward prediction error of RL for $(i, k)$ module

In motor-related cortices (Fig. 2b), computations similar to those in sensory-related cortices are executed utilizing sensory-motor conjugate-model pairs. Ultimately, the mismatch error and the reward prediction error are computed and weighted for summation, as in sensory cortical areas. The most marked differences between the sensory and motor cortices are directions of input-output of forward and inverse models. We set the higher level of the hierarchy for both sensory and motor cortices with more abstract representations, and the higher level is closer to the PFC. The lower level of hierarchy is characterized with rawer representations, and is closer to sensors, muscles, and the body. The inverse model (controller: Inverse: $f$) in motor cortices computes the motor command from the desired state, so its input-output direction is from top to bottom, which is the reverse of those of sensory inverse models. The important assumption is that the motor command of the higher level $u_H$ is identical to the desired state of the lower level $x_{dL}$.

$x_{dL} = u_H = f_H(x_{dH}, \hat{x}_H, \hat{x}_L)$ (7)

$x_{dLL} = u_L = f_L(x_{dL}, \hat{x}_L, \hat{x}_{LL})$ (8)

The forward model (state predictor: Forward: $g$) in sensory-motor cortices predicts the state from the motor command and the current state, so its input-output direction is from bottom to top, which is the reverse of sensory forward models.

$\hat{x}_H = g_H(\hat{x}_L, u_H)$ (9)

$\hat{x}_L = g_L(\hat{x}_{LL}, u_L)$ (10)

The mismatch of conjugate models is the difference of the desired state computed by the inverse model and the estimated state computed by the forward model.

$\| x_{dH} - \hat{x}_H \|$ (11)

$\| x_{dL} - \hat{x}_L \|$ (12)

The mismatch is expected to be small, provided that the module is relevant for current actions, that the inverse model is appropriate for motor control, and that prediction by the forward model is accurate. If the task requires eye movement, while the module is about foot movement, the mismatch should be large.

Again, the reward prediction error $\delta_{ik}$ is computed for the representation, and is weighted by a weighting factor $w$ and summed with the above mismatch. Here, $i$ represents the modality, such as vision, audition, or somatosenses. $k$ represents the hierarchical level, and corresponds to *LL, L, H, HH* in previous notations, and takes larger values for higher levels of the hierarchy. $\delta_{ik}$ is the difference between the predicted reward utilizing the $(i,k)$ representation and the actual reward.

The weighted summation will be used by CRMN. We call this weighted summation of the two kinds of errors "cognitive prediction error" $e_{ik}$. Eqs. 13 and 14 are for motor and sensory conjugate pairs, respectively.

$e_{ik}^2 = \| x_{ik}^d - \hat{x}_{ik} \|^2 + w\delta_{ik}^2$ (13)

$e_{ik}^2 = \| x_{ik} - \hat{x}_{ik} \|^2 + w\delta_{ik}^2$ (14)

The cognitive prediction error is not simply a sensory prediction error, a motor command error, or a reward prediction error, but contains all three types of errors. That is why we use the term "cognition" for this error.

The whole architecture is basically hierarchical reinforcement learning with multiple modules. One of the new features is that conjugate-model pairs constitute each hierarchy of each modality. Another feature is gating by CRMN, as explained below. The likelihood ($L_{ik}$: *likelihood of* $i - th$ *parallel module and* $k - th$ *hierarchy*) is a soft max function of the cognitive error.

$L_{ik} = \frac{exp(-e_{ik}^2/\sigma^2)}{\sum_{lm} exp(-e_{lm}^2/\sigma^2)}$ (15)

$\sum_{ik} L_{ik} = 1$ (16)

The responsibility signal $\lambda_{ik}$ is equal to $L_{ik}$ for small and intermediate $k$, but is defined as the product of the likelihood and a prior estimate of the responsibility signal, based on representations in all modules at high levels of the hierarchy.

$\lambda_{ik} = \hat{\lambda}_{ik} \cdot L_{ik}, k >> 1$ (17)

The terminology "responsibility signal" was borrowed from the MoEXP and MOSAIC literature. For a given task and state of the world, there may exist a well-suited module to cope with them. The gating network and CRMN select an appropriate module which is "responsible" for the task and the state, and computes the degree of this appropriateness by the responsibility signal. If it is large, the module is more appropriate and more heavily recruited for perception, action, and learning.

The prior estimate of the responsibility signal $\hat{\lambda}_{ik}$ roughly corresponds to the gating network of MoEXP architecture, as well as a discriminator network of GAN. $\hat{\lambda}_{ik}$ could be initialized as a flat vector or with random numbers. The function $h$ in Eq. 18 is incrementally updated (learned) with the $\lambda_{ik}$ as a teaching signal.

$$\hat{\lambda}_{ik} = h(x_{il}^d, \hat{x}_{il}, x_{il}), l \gg 1, k \gg 1 \qquad (18)$$

The responsibility signal $\lambda_{ik}$ uses cognitive error signals of all levels of the hierarchy for its computation, but its prior $\hat{\lambda}_{ik}$ does not use cognitive error signals, and instead can be estimated from abstract information sent by feedforward pathways, so by inverse models in sensory streams and forward models in motor streams, as well as abstract information broadcasted away by feedback pathways, so by forward models in sensory streams and inverse models in motor streams. Only high-level representations are used for the prior, as indicated by $l \gg 1$. Without depending on cognitive error signals, just by examining dimension-reduced abstract representations, CRMN can determine that some modules are most likely appropriate for representing the sensory world and the executed action, but that other modules are inappropriate. Based on the responsibility signal, the PFC selects the best conjugate-model pair for motor control and perception. This is a computational account for metacognition in the sense that the PFC "attends" to and "adopts" the selected module and hierarchy. The responsibility signal is used for perceptual attention and action selection, as well as gating learning of corresponding conjugate-models and reinforcement learning. The responsibility signal also acts as a teaching signal for its prior estimate. The CRMN in the PFC consists of the prior estimate network for the responsibility signal, a soft max to compute the responsibility signal, and computation of entropy $S$ of the responsibility signals as follows:

$$S = -\Sigma \lambda_{ik} \ln(\lambda_{ik}) \qquad (19)$$

Because information collected by CRMN is a very small subset of all representations in all modules and hierarchy, CRMN reduces data dimensions, and conducts module and feature selection by the responsibility signal. If we graphically expand feedforward and feedback pathways of conjugate model pairs in the right and left sides to the PFC, the whole neural network looks like an autoencoder network (hourglass model) and CRMN corresponds to its bottleneck with the smallest intermediate layer (Fig. 3b).

The most novel proposal for CRMN is its metacognition and consciousness explanations in addition to selection of modules for action, perception, and module-learning, as well as reinforcement learning. This is motivated by the experimental findings of the negative correlation between the decoded confidence and the decoded reward-prediction error in Cortese et al. (2020). We postulate that consciousness is determined by entropy $S$ of responsibility signals (Eq. 19), which is compatible with a previous coherence proposal of consciousness (Kawato 1997). If all responsibility signals are similarly small and almost uniform, entropy is large, and neither metacognition nor consciousness emerges. If one responsibility signal is much larger than other responsibility signals, then entropy becomes small, and metacognition of the cognitive process, which is executed by the module with the largest responsibility signal, emerges. The participant becomes conscious of corresponding perception and action representations. The function of the discriminator in GAN in (Gershman 2019) is replaced by the responsibility-signal entropy in CRMN. GAN's discriminator helps to train GAN's generator in a self-supervised fashion (Goodfellow et al., 2014). Someone first needs to train the GAN's discriminator with real data and false data. Separation of the data into the two classes necessitates either human involvement, or other brain parts, or homunculus. In the GAN algorithm, a teaching signal for discriminating between real and false data is necessary (here we note the presence of training data in GAN as the supervised learning for clarity), but we cannot afford to assume such a luxury in biological brains. If we make such an assumption, it is almost identical to assuming the homunculus who tells some neural activities representing the external world and others not. Instead of the supervised learning paradigm of GAN, reinforcement learning based on rewards and miniimization of responsibility-signal entropy are driving forces for all kinds of learning in CRMN architecture. This learning involves conjugate-model pairs, reinforcement learning, and the responsibility-signal prior estimator. For rigorous computation of the responsibility signals, the mismatch at each hierarchy and module should be computed. The responsibility signals can be approximated by the responsibility estimator that uses only the high-level top-down and bottom-up signals. Thus, in CRMN, the cognitive error signals are "teaching" signals in supervised learning, and the responsibility estimators are learned with this supervision. In this sense, CRMN was inspired by Gershman's proposal of GAN discriminator for reality monitoring and turned it into a self-supervised learning framework (Goodfellow et al., 2014; Gershman 2019; Lau 2019).

## (5) Relationships to previous models

CRMN is based on several lines of previous artificial neural network models and computational neuroscience models. In this section, we discuss how the proposed metacognition model explains various enigmas associated with learning from small samples in relation to previous studies. First of all, the PFC in CRMN serves as the bottleneck layer of the autoencoder neural network in dimension reduction, as explained in the previous section (Fig. 3b). Dimensional reduction in the PFC and selection of a module are the main mechanisms enabling learning from small samples. Because the huge original problem is transformed into many small and tractable problems with reduced dimensions, the PFC and the selected module can learn from a small sample.

     Conjugate-model pairs were proposed as forward and inverse models for cerebellar internal models (Kawato et al., 1987), as well as in MOSAIC (Wolpert and Kawato, 1998; Haruno et al., 2001), and proposed as essential elements in the coherence model of consciousness (Kawato 1997). In the perceptual domain, Kawato and colleagues (1993) proposed forward and inverse optics models for fast visual computation. The MoEXP model incorporated competition and cooperation between multiple modules in selection and learning (Jacobs et al., 1991). CRMN extends these previous models and incorporates generative-inverse model pairs as well as competition and cooperation between them, and is more general in the sense that perception and motor control are coherently managed. The novel feature of CRMN is its selection and gating mechanism. In previous models, a module with better learning performance (MoEXP), or a module containing a forward-model with better sensory prediction (MOSAIC) is selected. Thus, gating requires comparison of each module's output and teaching signal (MoEXP) or the state of the external world (MOSAIC). This may be possible within a shallow hierarchy, but would be practically impossible in brains with deep hierarchies. CRMN instead utilizes consistency of conjugate-model-pair predictions for gating, which can be evaluated within each level of hierarchy without reference to the global teaching signal or the sensory inputs. Our basic assumption is that the cascade of a generative model and its inverse should be an identity function; thus, the mismatch between corresponding outputs should be zero if the model pair is suited for a given context and is perfectly accurate (Fig. 2).

     Furthermore, CRMN is also based on a second stream of theoretical and computational studies: hierarchical and modular reinforcement learning. CRMN proposes that reinforcement learning occurs in parallel while utilizing individual representations for each module in the

hierarchy with its specific conjugate-model pair. Therefore, if the brain contains 10,000 different representations, and 10,000 different value functions, policies and reward prediction errors are estimated simultaneously and in parallel by loop computations between the basal ganglia and each module in the hierarchy (Fig. 3a). Module selection in CRMN is based not only on the consistency between conjugate-model pair computations, but also appropriateness of its representation for reinforcement learning. Thus, CRMN selects and recognizes a cognitive process not only by its perceptual consistency or motor-control competence, but also based on its optimality in maximizing long-term rewards. CRMN is closely related to RL-MOSAIC (Doya et al., 2002; Sugimoto et al., 2012a, 2012b) in the hierarchical-reinforcement learning literature in the sense that prediction goodness selects the appropriate module. While only goodness of forward-model prediction is considered in RL-MOSAIC, consistency of forward and inverse models is a criterion for selection, and is generalized to both perceptual and sensory motor domains in CRMN. CRMN also incorporates the reward prediction error in selecting an appropriate module, and this is a new feature, compared with MoEXP, MOSAIC, and RL-MOSAIC.

Computation of responsibility signals from cognitive prediction errors transpires in the PFC, and is the most novel feature of CRMN. This computation is closely related to the gating functions of MoEXP architecture, and the responsibility signal of MOSAIC. Estimation of the responsibility prior is related to prior estimation in MoEXP, and the discriminator network of GAN. If entropy of the responsibility signals is large, CRMN believes that all modules are inappropriate for perceptual interpretation of the current world, or for behavioral adequacy in a given task. This corresponds to a state in which the discriminator of GAN believes that the input is artificial. The small entropy of CRMN corresponds to a state in which the discriminator of GAN believes that the input is real. CRMN, as a model of metacognition and consciousness, is highly motivated by theories of Lau (2019) and Gershman (2019). A new aspect of CRMN is entropy of responsibility signals based on model mismatch and reward prediction errors, which can theoretically abolish the necessity of teaching signals for the GAN discriminator. Gershman (2019) explicitly proposed the prefrontal discriminator of GAN, which is a centralized consciousness model (higher-order theory). Our model is half centralized as responsibility signals and its predictor converge in the PFC, but half distributed (first-order theory) because computations of mismatches are individually executed at each hierarchy and module all over the cerebral cortices. The feedback generator and feedforward generator of Figure 1 in Gershman (2019) correspond to our forward optics (generative) model and inverse optics (inference) model (Fig. 2), respectively. The usage of "forward" and "feedback" are just opposite in both, and we note this just to avoid possible confusion.

Because CRMN is based on hierarchical-modular reinforcement learning, MOSAIC, and MoEXP, its origin lies in optimal action selection, and supervised and reinforcement learning. Thus, its ancestry had nothing to do with phenomenal consciousness or metacognition. But once CRMN is laid out, it could be a model of phenomenal consciousness, even though not a model of accessibility or access consciousness or attention. We propose that only when responsibility signal entropy $S$ is low, participants become consciously aware of representations with high responsibility signals. This is because all representations at several hierarchical levels are coherent, consistent and stable; thus, the PFC abstract representations are very well connected to first-order representations.

## (6) DecNef experimental support for CRMN

In this section, we discuss how the proposed metacognition model explains experimental results of decoded neurofeedback (DecNef), especially experimental results of Cortese and colleagues (2020). In the CRMN model, the search space of the reinforcement-learning state for any new task starts from all brain areas, and proceeds to very limited regions selected by CRMN, based on responsibility signals. This accords well with the experimental finding that brain areas correlated with the reward prediction error were widespread and covered the entire brain on the first day, but shrank very quickly to a few areas, including the basal ganglia and the PFC during the following two days. Participants with higher metacognitive capability learned faster in the reinforcement-learning task. Moreover, better actions were taken when participants were more confident about their perceptual judgments. If CRMN performs well, both metacognition and gating by the responsibility signal are efficient. Thus, the association between metacognition and learning found in the experiment is compatible with CRMN. Interestingly, in this study, participants could discriminate the correctness of their brain state inference (perceptual choice), but they were not conscious of the brain state itself. For the CRMN, to be conscious we need to have a high responsibility signal across the distribution of representations that span the entire hierarchy – lower-level representations must match higher-level ones. Nonetheless, for metacognition the responsibility signal is about a single module; thus, even with a relatively low, but graded responsibility signal, this could be sufficient to have metacognitive insight, while failing to reach consciousness. As learning progressed, confidence decoded from PFC multi-voxel patterns showed larger information coupling with reward prediction errors decoded from basal-ganglia multi-voxel patterns. CRMN performs several functions simultaneously and coherently, based on responsibility signals. Functions include gating modules for metacognition, perception, and action, and both conjugate-model learning and reinforcement learning. Thus, experimental results are compatible with predictions from CRMN.

CRMN can explain the "consciousness enigma" of DecNef experiments (Shibata et al., 2011; Cortese et al., 2016, 2021a). Although decoded fMRI neurofeedback in these experiments induced strong brain representations and caused significant behavioral changes with medium to large effect sizes (Watanabe et al., 2017), participants remained unaware of the information induced in their own brains (Shibata et al., 2018; Cortese et al., 2021a). CRMN predicts that strong neural representations in a module and hierarchy are not sufficient to induce consciousness. Feedforward and feedback computations by conjugate-model pairs should be compatible as a prerequisite for consciousness. However, the target area of DecNef was isolated from other areas regarding viewpoints of the induced information (Shibata et al., 2011, Amano et al., 2016). Furthermore, although the induced information was strong enough to change related behaviors, no relevant sensory stimulus or motor task was given to participants; thus, neither generative models nor inverse models receive appropriate inputs related to the induced information from lower and/or higher levels of the hierarchy. Without relevant inputs, neither model can accurately predict information representations in the targeted area in DecNef. Consequently, mismatches between the conjugate model-pair should be quite large and the corresponding responsibility signal should be small. In this case, CRMN asserts that the information representation in that area does not reach the level of consciousness, even though it is strong enough to change behavior. Consequently, CRMN explains why participants in DecNef did not become conscious of the information induced in their own brains that was strong enough to cause behavioral changes.

## (7) Relationship to phenomenal consciousness

In this section, we discuss how the proposed metacognition model explains experimental results of a broad array of phenomena in metacognition, consciousness, and sense of agency. On the same basis of larger mismatch signals introduced in the previous section, CRMN explains why spontaneous brain activities (Berkes et al., 2011; Kenet et al., 2003; Luczak et al., 2009), brain activities induced by working memory, or brain activities induced by mental imagery, are not necessarily brought above consciousness. Brain activities under these conditions may contain strong information representations in some brain areas (Albers et al., 2013; Mendoza-Halliday and Martinez-Trujillo, 2016), but these areas lack either feedforward or feedback neural inputs or both, regarding relevant representations from lower and higher levels in the hierarchy. Thus, the mismatch within the conjugate-model pair is large and responsibility signals of the areas are small. During working memory and mental sensory imagery, top-down signals are sent from the PFC to sensory cortices, but bottom-up signals are absent because of the lack of corresponding sensory stimuli. Therefore, mismatch signals should be large at many levels of the hierarchy, and especially at lower levels. In the CRMN framework, large mismatch signals imply nonconscious representations. Nevertheless, persistent local recurrence between generative and inverse models at some intermediate levels may generate internal signals travelling upward to middle and high levels of the hierarchy, leading to small mismatch with top-down signals (conscious representations). These cases may correspond to conscious experiences for some participants. Thus, CRMN predicts that the efficiency of recurrent computations between the forward and inverse models and their levels in the hierarchy determines how strongly a participant is conscious about working memory or mental imagery content.

Top-down signals sent from the PFC through feedback pathways in the sensory stream and top-down pathways in the motor stream cannot tell whether downstream representations are real or not. Let us take a thought experiment. The motor center sends a rightward finger movement intention, but actually a participant is involved in a sense of agency experiment, and is shown a leftward finger movement video. Accordingly, sensory representations of the rightward movement inferred from the motor command are not real. We can think of similar mismatches between working memory and visual imagery top-down signals and real bottom-up sensory signals. An important proposal of CRMN is that only if the brain examines compatibility of top-down and bottom-up signals, can it discriminate real from spurious representations of the external world and executed task.

The peculiar cases of aphantasia and blindsight offer important tests for the CRMN model. People with aphantasia can perform mental imagery, but lack conscious awareness of the imagery content (Pounder et al., 2021). In blindsight, people are unable to consciously experience visual stimuli due to damage to the visual cortex. Yet, they are able to objectively discriminate between visual stimuli (Weiskrantz 1996; Stoerig, Cowey 1997). CRMN can accommodate both conditions. In aphantasia, the lack of consciousness may result from weaker recurrent local interactions between the generative model and its inverse model, or deficits in the CRMN. In blindsight, bottom-up signals are weak; thus, high-level representations are not formed, and top-down signals are absent, so mismatch signals are large. However, the weak low-level signals may still be used to perform tasks that involve unconscious processing, such as forced choices, Pavlovian and instrumental conditioning, reinforcement learning (Hamm et al., 2003; Takakuwa et al., 2017; Kato et al., 2021).

Based on similar reasoning, CRMN provides an explicit computational account of "sense of agency" (Haggard 2017) and its deterioration in brain disorders (Biran, Chatterjee 2004; Fried et al., 2017). Mirror neurons of monkeys and the mirror-neuron system of humans are activated both in movement control and observations of related movements executed by other agents (Di Pellegrino et al., 1992; Iacoboni et al., 1999; Rizzolatti 2004). Wolpert and colleagues proposed that forward models in MOSAIC can be utilized to infer intentions of other agents from observations of their movements (Wolpert et al., 2003). Even though neural representations of one's own movements and movements executed by others could be similar in motor related cortices, we can well discriminate the two types, i.e., we have capability of sense of agency. Who executes the movements observed by the sensory system? Myself or someone else? CRMN explains the difference between these two cases by inverse models and top-down neural connections. When one visually observes a video of arm movements, for example, bottom-up neural connections carrying visual information and resulting forward model computations could be similar to those generated by one's own movements. However, because there is no movement intention in higher motor cortical areas, top-down neural connections do not carry much control input to inverse models; thus, the mismatch between the inverse models and forward models becomes large. Then, CRMN rejects the hypothesis that neural representations in premotor or motor cortices are generated by one's own movements, and the sense of agency does not emerge. In visual cortical areas, the mismatch could be small and participants may be consciously aware of arm movements. This capability is compromised in patients with psychosis (Blakemore et al., 2000), and we can explain this either by deficits in CRMN, mismatch computation, or within forward (Blakemore et al., 1998) or inverse models. Conscious perception is also similarly compromised in schizophrenia (Berkovitch et al., 2017). Finally, based on similar CRMN mechanisms, brain activities induced by mental motor imagery and spontaneous brain activities are rejected for representing actual movements, even if they have similar motor representations in some brain areas. Because CRMN is based on a single set of computations, it can offer a simple unifying explanation for the seemingly different phenomena of psychosis, schizophrenia, conscious perception and agency.

We assume that, in the PFC/CRMN, an abstract representation of conscious content exists (regardless of whether it is $L_{ik}$s or vectorial representations), and only if this content is compatible with first-order content in the sense of high responsibility signals. Then conscious awareness of the content emerges. In this paper, we did not consider several interesting phenomena in conscious vision, including peripheral inflation, binocular rivalry, masking, weak stimuli, but most of these may be explained by the interplay (mismatch between the conjugate pair or coherence between first-order and higher-order representations) between generative and inference models and their monitoring (in the way of more abstract representations and responsibility signals) by the PFC. It is one of our future research plans to systematically consider these interesting cases from the theoretical framework of CRMN.

## (8) Discussion

Here, we discuss the proposed model of metacognition in the context of several previous theories of consciousness. We further speculate how this computational neuroscience model could lead to next-generation artificial intelligence, which possesses metacognition and consciousness (Dehaene et al., 2017), and is able to learn gigantic problems from small samples. To conclude, we illustrate how this model, at the intersection of neuroscience and artificial intelligence, may also inspire new experiments to causally induce consciousness in humans.

CRMN provides an implementable computational algorithm for higher-order theories (HOT) of consciousness (Brown et al., 2019). In general HOT, the PFC monitors a cognitive process as a requisite of metacognition. However, it has not been computationally specified what kinds of neural information are monitored. It is also not evident whether this monitoring mechanism is fundamentally different from a homunculus observing a cognitive process. In CRMN, the PFC monitors coarse-grained information representations, mismatch signals, and reward prediction errors from each area. Of note, while monitoring is about coarse-grained information, what the agent will be consciously aware of are representations associated with small entropy, which can correspond to coarse- or fine-grained information. Thus, the PFC monitors only very limited information and computes responsibility signals in an algorithmic way, so it cannot be a homunculus. In CRMN, we explicitly propose that consciousness has important functional contributions to survival and learning from small samples. CRMN clarifies the relationship between metacognition and consciousness with a concrete algorithm. Both result from one underlying mechanism, but metacognition is defined over one cognitive process, and consciousness is defined over distributed variables (responsibility signals) of all cognitive processes and its entropy. As long as responsibility signals are large, the PFC can simultaneously accommodate metacognition of multiple cognitive processes. Because consciousness is determined by the entropy of all responsibility signals, it is a state determined by all modules, and different from metacognition.

CRMN is also consistent with a second major hypothesis of consciousness in some regards. The prerequisite of consciousness in CRMN is a low mismatch, leading to a high responsibility signal between a generative model and its inverse model. Such a situation can only be attained if the two paired models are consistent and signals circulating between them are self-consistently maintained in that brain area. This accords with the basic assumption of recurrent processing theory (Lamme et al., 2000; Lamme 2018, Malach 2021). Recurrent neural connections within and between areas are essential to maintain neural activities above consciousness. One-shot feedforward computations across hierarchical brain areas are not sufficient for consciousness in the recurrent processing theory or in CRMN. With this note, one may notice that CRMN could not be perfectly classified as a higher-order theory because the computations of mismatch signals and reward prediction errors, both of which are at the heart of the responsibility signal computation, are done in a distributed manner by every level of the hierarchy and modality, and by their loop connections with the basal ganglia. In that sense, CRMN may be regarded as a blend of first-order theory and higher-order theory. As prerequisites of consciousness, CRMN requires two conditions: that a first-order representation of some module and some level in the hierarchy have a small mismatch, and that the PFC detects it by a small entropy of the responsibility signals.

In CRMN, the PFC is an information hub. Information from all cerebral cortical areas is collected in the PFC, which sends responsibility signals back to all cerebral cortical areas in return. Thus, communications between the PFC and all cerebral cortical areas are essential for consciousness to be established in CRMN, as in global neuronal workspace theory, another major theory of consciousness (Dehaene and Naccache 2001). Yet, a delicate difference in nuance of information broadcasts may exist. Widespread information broadcasts between several areas of key cognitive functions are not essential in CRMN, although they might be in global neuronal workspace theory.

One of the core proposals of the self-organizing mental representational account (SOMA) theory by Cleeremans et al. (2019) is that learning of a metarepresentation of first-order representations is essential for consciousness. If we can assume that higher level representations with generative and inference model pairs at higher and more abstract levels of CRMN roughly correspond to metarepresentation of SOMA, the two theories are related.

Currently CRMN does not explicitly include the default mode network (DMN). We have the following hypothesis regarding partition of labor in reinforcement learning (RL) between task positive networks versus task negative networks or DMN. The former is mainly for on-line currently experienced trials for RL, while the latter is for off-line, off-policy, mental simulations of RL. The responsibility signals and learning rates of on-line trials should be generally larger than those of off-line trials. Furthermore, different values of other RL hyperparameters should be chosen for on-line and off-line RL trials. Thus, the ventromedial PFC and the dorsolateral PFC may contain different architectures of CRMN for selection and gating of modules in DMN and the task-positive network, respectively. The DMN is heavily involved in social cognitive functions; thus, it seems to be involved especially in off-line social simulations of RL. Wolpert et al. (2003) proposed that conjugate model pairs are utilized to understand the intention of actions by others. In this context, the expanded CRMN seems related to a proposal by Graziano and Kastner (2011) and Fleming (2020) that the same mechanisms/brain areas involved in attributing agency to others are potentially involved in generation of consciousness in a given agent. Further, multiple accounts of metacognition have modeled self-reflection as an inference about others' performance or mental states (Fleming, Daw 2017; Shea et al., 2014).

Bengio (2017) noted that language is important in the consciousness prior for artificial intelligence. Language allows complex sensory information to be represented at a symbolic level. Previous human imaging studies suggest continuous transitions from sensory-motor domains to language. We demonstrated that the human cerebellum contains multiple internal models of tools (Imamizu et al., 2000, 2003; Higuchi et al., 2007), and also showed that representations of language and tools overlap in Broca's area, supporting the tool-origin theory of language (Higuchi et al., 2009). Beyond dimensionality reduction and selection of the best modules or representations, the architecture proposed here would endow an artificial agent with an additional advantage. A scheme that updates multiple internal models in parallel for each acquired data point would enable the agent to modify its internal states even in modalities that are not immediately relevant. This seems particularly important in the real world, where relevant data are often sparse and delayed, and environments are stochastic. What is unnecessary now might be crucial an hour later. The CRMN implements two aspects of human reasoning: hypothesis testing that probably happens at the conscious level, but also keeping track of multiple alternatives subconsciously. Engineering implementations of generative models, their inverses, and estimators of responsibility signals with deep neural networks may inaugurate new-generation artificial intelligence with metacognition, consciousness, and learning capability from small samples. As a small step in this direction, we plan to computationally simulate a simplified CRMN as a MoEXP architecture. To this end, we will model participant learning behaviors observed in Cortese et al., (2021b). In that experiment, participants were able to learn an abstract representation for reinforcement learning from a small sample. The fact that participants' confidence was significantly correlated with their ability to select the correct abstraction is a strong indicator of its involvement in accelerating learning. Our ultimate, and possibly long-term scientific goal is to cause phenomenal consciousness by decoded neurofeedback without sensory stimuli or motor tasks. Knotts et al. (2019) obtained intriguing results toward this direction, finding that reinforcing mental representations of high confidence and a stimulus feature (e.g., the color red) was associated with a higher chance of making false alarms. Yet, this study also showed that the goal of generating phenomenal consciousness is probably too difficult without explicit computational models like CRMN. With multi-voxel decoding of responsibility signals based on CRMN, a "causal study of consciousness" may be within our reach.